\documentclass[12pt,reprint,aps,groupedaddress,amsmath,amssymb,onecolumn,]{revtex4-2}

\usepackage{hyperref}  
\usepackage{xr-hyper}

\usepackage{subcaption}

\usepackage{float}
\usepackage{braket}
\usepackage{graphicx}
\usepackage{dcolumn}
\usepackage{bm}
\usepackage{geometry} 
\usepackage{newtxtext,newtxmath} 
\usepackage[version=4]{mhchem} 
\usepackage{upgreek} 
\usepackage{rotating}

\usepackage{ragged2e}  
\DeclareCaptionJustification{justified}{\justifying}

\usepackage[]{caption}
\usepackage{stackengine}
\newcommand{\eex}{\stackengine{0.4ex}{\^e}{\~}{O}{c}{F}{T}{L}}

\newcommand{\cmmnt}[1]{\ignorespaces}
\begin{document}

\title{Scanning Tunneling Microscope Tip-Induced Formation of Bi Bilayers on Bi$_2$Te$_3$}

\author{Duy Nguy\eex{}n}
 \email{nguyen.2502@osu.edu}
\author{Jay A. Gupta}
 \email{gupta.208@osu.edu}
\affiliation{Department of Physics, The Ohio State University}
\date{July 21, 2025} 

\begin{abstract}
We report the formation of Bi(111) bilayer (BL) islands and crater structures on Bi$_2$Te$_3$(111) surfaces induced by voltage pulses from an STM tip. Pulses above a threshold voltage ($+3$ V) produce craters $\sim 0.5$ microns in diameter, similar to the size of the tip. Redeposited material self-assembles into a network of atomically ordered islands with a lattice constant identical to the underlying Bi$_2$Te$_3$ surface. The island size monotonically decreases over several microns from the pulse site, until the pristine Bi$_2$Te$_3$ surface is recovered. We assign these islands to Bi BL based on atomic resolution images, analysis of step heights, and tunneling spectroscopy. The dependence of bilayer formation on bias polarity and the evidence for defect diffusion together suggest a mechanism driven by the interplay of field evaporation and tunneling-current-induced Joule heating.
\end{abstract}
\maketitle

\textbf{\textit{Introduction.}} The prediction of the quantum spin Hall (QSH) phase in single-bilayer (BL) Bi(111) has sparked significant interest in exploring BL structures on three-dimensional topological insulators (3D TIs), such as the Bi\textsubscript{2}(Te, Se)\textsubscript{3} family \cite{predicnultrathinBi, 1stBionBi2Te3,2ndBionBi2Te3,arpesBionBi2TeSe3}. These systems, where a two-dimensional TI interfaces with a 3D TI, have prompted intensive investigations into emergent topological phenomena \cite{arpesBionBi2TeSe3, QAHE, topologicalsuperconductor}. Bi BLs on Bi-based TIs have been achieved through various approaches: molecular beam epitaxy (MBE) \cite{2ndBionBi2Te3, BionBi2Te2Se}, sputtering and annealing treatments on Bi$_2$Te$_3$ surfaces \cite{BiannsputBi2Te3, BiannBi2Te3}, and spontaneous formation upon cleaving Bi$_2$Se$_3$ at room temperature \cite{BionBi2Se3, BionBi2Se32}. To achieve localized nanoscale patterning, a previous study applied voltage pulses via the STM tip to reproducibly pattern bulk Bi surfaces  \cite{Bi111pulse}. The restructuring mechanism was attributed to electric field evaporation, supported by the observed linear dependence of a threshold voltage on tip–sample separation.

Here we demonstrate an STM-based method that can form Bi BL nanostructures on Bi$_2$Te$_3$, with preferential removal of Te during the process. Positive voltage pulses ($\ge 3$ V) produce a crater at the pulse site with a surrounding network of Bi BLs. In contrast, negative voltage pulses produce a similar crater but do not produce the Bi BL structures. Atomic resolution imaging indicates the BL is lattice-matched to the underlying Bi$_2$Te$_3$. An inhomogeneous distribution of point defects in the vicinity of the crater, and the observed threshold and polarity dependence, suggest an underlying mechanism sensitive to electric field evaporation and Joule heating.

\textbf{\textit{Methodology.}} All STM measurements were conducted at  $4.5 \, \text{K}$ in an ultrahigh vacuum (UHV) system ($10^{-11} \, \text{mbar}$ ) using a CreaTec LT-STM system. Mechanically cut PtIr and electrochemically etched Ni, Cr, and W tips were used in the experiments, with apices characterized by scanning electron microscopy (SEM). Unless otherwise noted, all data presented here were taken with the Ni tip. Scanning tunneling spectra ($dI/dV$) were acquired via lock-in detection by adding an AC bias modulation of $50$ mV at $877$Hz. Topographic images were processed with the Gwyddion software package \cite{Necas2012}. The Bi$_2$Te$_3$ single crystal was purchased from 2D Semiconductors. To expose a clean surface, a small metal post was attached to the crystal surface using UHV-compatible EPO-TEK\textsuperscript{\tiny\textregistered} H21D epoxy, cured at $150^\circ$C for 1 hour. The sample was then transferred into the UHV chamber at room temperature, where the post was mechanically knocked off using a translator to expose a fresh Bi$_2$Te$_3$(111) surface. The sample was then quickly transferred into the cold STM.

\begin{figure}[h!] \centering 
\includegraphics[width=1\textwidth]{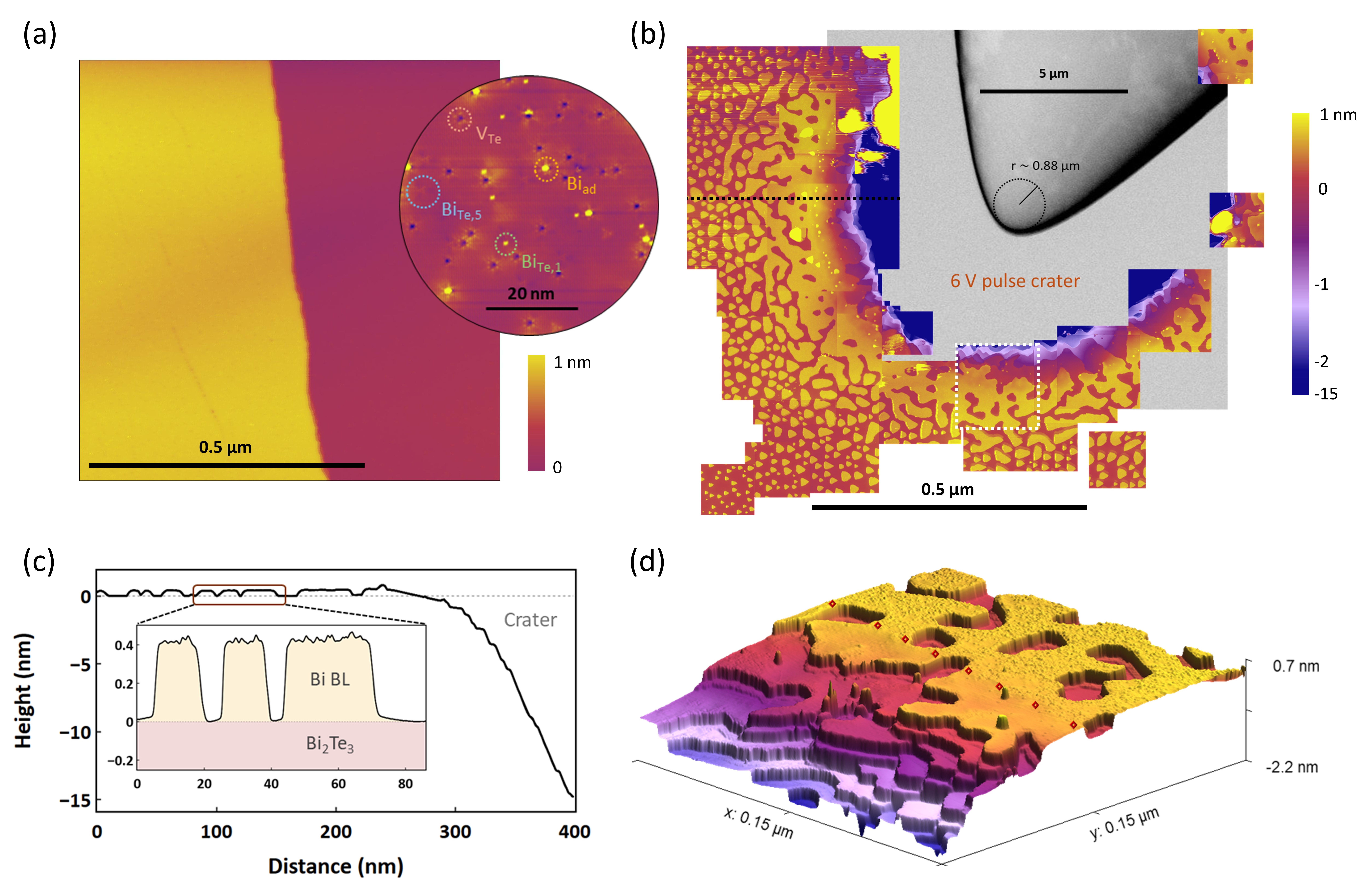} 
\caption{\textbf{STM topographic images after $6$ V tip pulse.} 
\textbf{(a)} Image of pristine Bi$_2$Te$_3$ showing a single QL step, $\sim 1\ nm$. \textbf{(Inset)} Atomic-resolution image showing point defects ($0.5$ V, $1$ nA). 
\textbf{(b)} Mosaic STM image of the area affected by a $6$ V tip pulse, capturing the pit crater and the formation of Bi BL islands on the Bi$_2$Te$_3$  terrace. \textbf{(Inset)} SEM image of the electrochemically etched Ni tip. 
\textbf{(c)} Height profile along the black dashed line in (b), showing the depth and sloping profile of the crater. \textbf{(Inset)} Zoomed-in height profile illustrating the well-ordered Bi bilayer structure with consistent step heights of approximately $\sim 0.39$ nm.
\textbf{(d)} Three-dimensional-rendered topographic image of the white dashed box in (b) showing the spatial distribution of defects near the crater. [Red diamonds] mark the distinct boundary separating the defect-rich and near defect-free regions on a Bi BL surface.} 
\label{fig1} 
\end{figure}

\textbf{\textit{Result \& Discussion.}} Figure~\ref{fig1}(a) shows an STM topographic image of the pristine Bi$_2$Te$_3$(111) surface. The surface typically exhibits micron-scale, atomically flat terraces with a low occurrence of atomic steps. These are usually $\sim 1$ nm in height, corresponding to a quintuple layer (QL) of the Bi$_2$Te$_3$ structure. Atomically resolved images (inset) show a triangular lattice attributed to surface Te atoms, and a low density of native point defects. The predominant defects, Te vacancies, appear as dark spots, and are consistent with previous reports \cite{nativeBi2Te3}. In acquiring such images, STM tip sharpening on Bi$_2$Te$_3$ is routinely performed by applying $1.5$ V - $2$ V pulses. Each pulse lasts for $\sim 200$ milliseconds with a fixed tip height of $\sim 1$ nm during the pulse and a current that reaches $\ge 10$ nA. 

We find that when higher voltage pulses are applied, the surface undergoes significant modification. Fig.~\ref{fig1}(b) shows an STM image after applying a $6$ V voltage pulse. A mosaic of topographic images stitched from multiple smaller-scale scans reveals a pit-like crater. The crater has a round shape of $\sim 0.4\, \mu$m in radius and reaches a depth of $\sim 70$ nm. While imaging at the crater bottom was typically unstable, the limited data suggest a flat morphology. The crater's size is comparable to the apex diameter of the etched Ni tip, as confirmed by SEM imaging  (Fig.~\ref{fig1}(b) inset). The line profile in Fig.~\ref{fig1}(c) and perspective image in Fig.~\ref{fig1}(d) show a complex staircase-like morphology along the edge of the crater, with a variety of sub-QL step heights observed. We attribute steps of approximately $0.39$ nm in apparent height to the Bi BL. Step heights distinct from the $0.39$ nm Bi BL step may instead correspond to different intermediate stacking configurations observed in the mixed-stoichiometry Bi$_m$Te$_n$ homologous series \cite{BiNTeM, BiNTeM2, Bi4Te3} (see Fig. SM4). For example, Bi$_2$Te$_2$ and Bi$_4$Te$_3$ have Bi BLs inserted between Bi$_2$Te$_3$ QLs at every second and first QL, respectively. A local sequence of three Bi$_2$Te$_3$ QLs adjacent to two Bi$_4$Te$_3$ stacks would result in a $\sim 0.2$ nm step height. These structural signatures suggest that the Bi BL formation process may extend below the surface to include subsurface reconstruction, potentially driven by Te depletion during the voltage pulse. 

Surrounding the crater is an atomically flat network with a dendritic morphology. With increasing distance from the pulse site, this network fragments into uniform, atomically flat islands that monotonically decrease in size. Utilizing the coarse positioning of the STM scan head, we traced the surface several microns away from the crater (see Fig. SM2), and confirmed that the original terrace is recovered. This allows us to define the original terrace height as the dashed baseline in Fig.~\ref{fig1}(c). This indicates that the islands and network are formed atop the existing Bi$_2$Te$_3$ terrace, from material excavated from the crater. Both the network and islands have a uniform step height of $0.39$ nm $\pm \ 0.01$ nm (Fig.~\ref{fig1}(c) inset), which we attribute to Bi BL regions. This is consistent with prior STM reports of Bi BLs on Bi$_2$Te$_3$ \cite{BiannsputBi2Te3, BiannBi2Te3, 1stBionBi2Te3, 2ndBionBi2Te3}. Figure~\ref{fig1}(d) also reveals an abrupt change in defect density on the Bi BL periphery, marked by red diamonds (see also Fig. SM3). When extended across the larger mosaic, this boundary is roughly circular and centered on the pulse site. Conversely, the crater edge itself appears irregular and jagged. These observations will be discussed further below in light of the underlying formation mechanism. 

\begin{figure}[h!]
\centering
\includegraphics[width=1\textwidth]{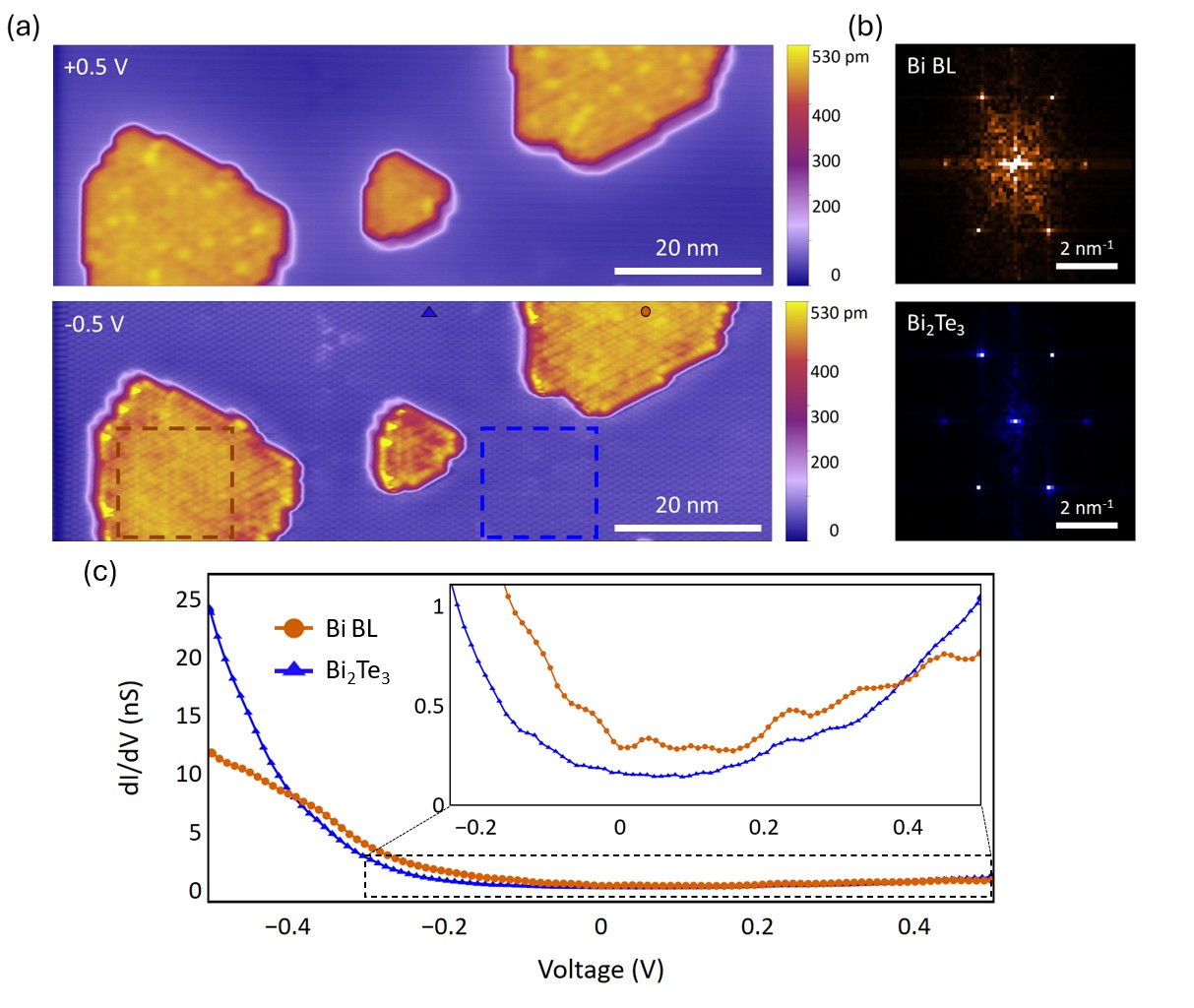}
\caption{\textbf{Electronic and structural characterization of Bi BL islands on Bi$_2$Te$_3$ terrace.} 
\textbf{(a)} STM topographic images of Bi BL islands acquired at sample biases of $+0.5$ V (top) and $-0.5$ V (bottom).  
\textbf{(b)} Fast Fourier Transform (FFT) images corresponding to the dashed boxes in (a), showing lattice constants of $0.44 \, \text{nm} \pm 0.07 \, \text{nm}$ and $0.45 \, \text{nm} \pm 0.06 \, \text{nm}$ for the Bi BL and Bi$_2$Te$_3$ surface, respectively. 
\textbf{(c)} Scanning tunneling spectroscopy ($dI/dV$) acquired on Bi$_2$Te$_3$ (blue triangle in (a)) and on a Bi BL island (orange circle in (a)). 
\textbf{(Inset)} Zoom-in of the low-bias region of the $dI/dV$ spectra.}
\label{fig2}
\end{figure}

To further confirm the Bi(111) BL, we examine the atomic structure of the islands in Fig.~\ref{fig2}. The topographic images in Fig.~\ref{fig2}(a) show that the islands appear defect-rich when imaged at $+0.5$ V, whereas a well-ordered periodic lattice structure becomes visible at $-0.5$ V. This contrast likely reflects an increased sensitivity to defect-related states in empty-state imaging, while filled-state imaging at negative bias more clearly reveals the underlying atomic ordering. Compared to the surrounding Bi$_2$Te$_3$, the BL islands exhibit a significantly higher density of point defects. The defect density is estimated to be $190 \pm 16$ defects per $1000$ nm$^2$, similar to the values reported for annealing-induced Bi BL on Bi$_2$Te$_3$~\cite{BiannBi2Te3}. In contrast, the Bi$_2$Te$_3$ surface surrounding the Bi BL island has a defect density of $4.7 \pm 0.4$ defects per $1000$ nm$^2$, which is notably lower than that of the pristine Bi$_2$Te$_3$ surface farther from the pulse site ($16.9 \pm 0.4$ defects per $1000$ nm$^2$). In particular, native defects commonly observed on freshly cleaved Bi$_2$Te$_3$—including V\textsubscript{Te}, Bi\textsubscript{Te,1}, and Bi\textsubscript{ad}  (see Ref.~\cite{nativeBi2Te3} for defect classification)—are absent up to several hundred nanometers from the crater. The remaining Bi$_2$Te$_3$ defects are primarily subsurface antisites, including Bi\textsubscript{Te,5} and Bi\textsubscript{Te,6}. This suggests that Te and Bi adatoms, displaced during pulse-induced surface disruption, may diffuse to passivate preexisting native defects in Bi$_2$Te$_3$ and contribute to the self-assembly of Bi bilayer islands. 

Fast Fourier Transforms (FFTs) of the BL and Bi$_2$Te$_3$ regions in Fig.~\ref{fig2}(b) show hexagons associated with a triangular lattice, with lattice constants of  $0.44\, \text{nm} \pm 0.07 \, \text{nm}$ and $0.45 \, \text{nm} \pm 0.06 \, \text{nm}$ respectively and complete rotational alignment. These data show that Bi BL is well-aligned with the underlying Bi$_2$Te$_3$ substrate, and is consistent with previous observations of epitaxial Bi BL grown by MBE or induced by annealing on Bi$_2$Te$_3$ \cite{1stBionBi2Te3, BiannBi2Te3}.

Figure~\ref{fig2}(c) compares the $dI/dV$ spectra acquired at the center of a Bi BL island (orange circle) with one from the surrounding Bi$_2$Te$_3$ terrace (blue triangle). The $dI/dV$ spectrum of Bi$_2$Te$_3$ exhibits a characteristic U‑shape. A slight bump near $+0.23$ V is assigned to the conduction band minimum (CBM) feature based on previous STM, angle-resolved photoemission spectroscopy, and density functional theory studies \cite{Bi2Te3, Bi2Te32}. The Bi BL spectrum has a broad minimum from $0$ V to $+0.2$ V but generally higher density of states between $-0.4$ V and $+0.4$ V compared to Bi$_2$Te$_3$. The overall $dI/dV$ spectral shape is generally consistent with previously reported STS of Bi BL on Bi$_2$Te$_3$, particularly Ref.\cite{1stBionBi2Te3}. However, we note that the broad minimum we observe is shifted relative to that in Ref.\cite{BiannBi2Te3}. This difference could be due to defect doping and/or quantum confinement states, which may be evident as weak peaks in ~\ref{fig2}(c) \cite{QWBiSb}. 

\begin{figure}[h!] 
\centering \includegraphics[width=1\textwidth]{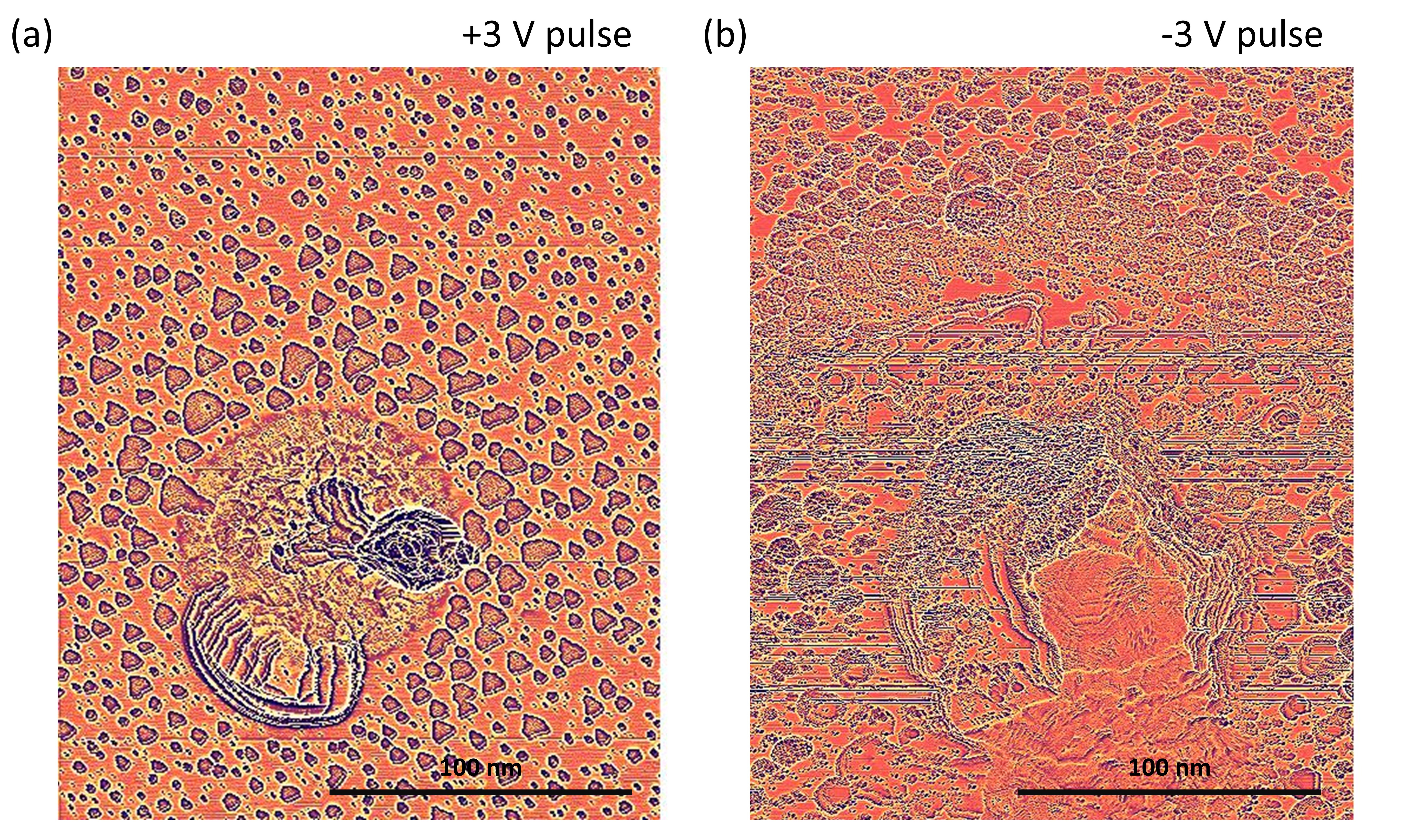} 
\caption{\textbf{Field-dependence of Bi BL formation} \textbf{(a)} STM image following a +3V pulse.  \textbf{(b)} STM image following a -3 V pulse. Both images were taken at $1$ V, $100$ pA with a Laplacian filter to emphasize local contrast} 
\label{fig3} 
\end{figure}

Turning now to the underlying mechanism, we first note that prior STM studies have proposed several mechanisms for tip-induced restructuring, including electric field evaporation, mechanical contact, local sublimation by tunneling electrons, and Joule heating \cite{tsong, gu, oyabu, skondo, staufer}. For example, Bi BL islands have been successfully extracted and re-deposited on bulk Bi films via STM tip pulsing \cite{Bi111pulse}. Electric field evaporation was identified as the dominant mechanism due to a threshold voltage that depended linearly on the tip–substrate separation. Similarly, the formation of Bi BL through annealing suggests that localized Joule heating may also play a role in the tip-induced process. \cite{BiannBi2Te3, BiannsputBi2Te3, BionBi2Te2Se, arpesBionBi2TeSe3}. To explore the mechanism behind crater formation and Bi BL island growth, we conducted a series of experiments with increasing pulse voltages, with the tip height fixed at $\sim 1$ nm under standard imaging conditions. We found that surface restructuring, including crater formation and Bi BL growth, only occurred for pulses above a threshold of $+3$ V. At lower pulse voltages, the surface remained unchanged, though in a few cases, small fragments or clusters appeared to eject from the tip and scatter around the pulse site.

Figure~\ref{fig3}(a) shows the Laplacian-filtered STM topography following a $+3$ V tip pulse. Compared to the higher voltage pulse in Fig.~\ref{fig1}, a smaller pit crater forms with a width of $\sim100$ nm and a depth of $\sim 9$ nm. Similar to the $+6$ V case, Bi BL islands form around the crater, though they are generally smaller and decrease in size with distance from the pulse site. In this instance, a mound appeared adjacent to the crater, suggesting material is falling from the tip apex after the pulse. The estimated volume of this mound, combined with the surrounding Bi BL islands, is on the order of $\sim 5 \times 10^4$ nm$^3$, which roughly matches the crater’s volume. This suggests that the Bi BL islands likely formed from Bi/Te atoms excavated from the Bi$_2$Te$_3$ substrate during crater formation. A smooth, round transition boundary can also be observed around the mound, illustrating the spatial extent of disturbance on the Bi$_2$Te$_3$ surface. Although not identical to the defect transition boundary seen in the $+6$ V case, both the pulse-induced craters exhibit a smoothly round transition boundary and jagged crater edge. 

In contrast, negative voltage pulses produce qualitatively different surface morphologies. As shown for a $-3$ V pulse in Fig.~\ref{fig3}(b), the resulting crater has a width of $\sim 200$ nm and a depth of $\sim 13$ nm, comparable to that formed by the $+3$ V pulse. The surrounding islands are flat but more irregular in shape, with a step height of $\sim 1$ nm near the crater that is similar to the Bi$_2$Te$_3$ QL, but distinct from the $0.39$ nm step height associated with the Bi BL. With increasing distance from the pulse site, the islands evolve into smaller clusters with rounded tops. These observations suggest that Bi BL structures are not generated under negative voltage pulses. 

The observed threshold voltage and polarity dependence suggest a field evaporation model. Cleaved Bi$_2$Te$_3$ surfaces are Te-terminated under UHV and up to room temperature \cite{Te_termination}. A positive voltage pulse ($\ge 3$V) produces an apex field on the order of $10$ V nm$^{-1}$, likely sufficient to exceed the field-evaporation threshold for Bi and Te \cite{SWANSON}. A similar process has been reported on bulk Bi surfaces, where field-induced evaporation produced craters above a threshold voltage of $\sim +2.5$ V under similar tunneling conditions \cite{Bi111pulse}. Here and as in prior studies, we observe that higher pulse voltages produce larger craters. We further find that the outcome depends on tip geometry. Etched Ni and Cr tips with apex radii around $1\, \mu$m reliably produce both craters and Bi BL islands.
In contrast, cut PtIr tips, which lack a well-defined sharp apex due to their fractured and irregular geometry, do not induce Bi BL formation. At the opposite extreme, an etched W tip with an apex radius $\le 50$ nm produced smaller craters ($\sim 40$ nm in diameter for a $+6$ V pulse), surrounded with displaced material and/or clusters rather than Bi BL structures. Both the voltage and geometry dependence are consistent with a field-driven mechanism.

In addition to field effects, local Joule heating from the tunneling current may also contribute to the restructuring process. During pulsing, the tunneling current exceeds the $10$ nA preamplifier limit, suggesting the possibility of $\ge 30$ nW transient heating. This localized heating could provide the thermal energy required for Te desorption, Te/Bi adatom migration, and surface reorganization. Analogous effects have been observed in Bi$_2$Te$_3$ under UHV annealing at temperatures up to $420^\circ$C, where surface Te desorbs and Bi segregates to form Bi BL structure \cite{BiannBi2Te3, BiannsputBi2Te3}. These studies also showed that annealing at elevated temperatures tends to favor Bi BL formation over mixed Te-Bi terminations. As mentioned earlier (Fig.~\ref{fig1}(d) and Fig.~\ref{fig1}(b)), we identified a boundary surrounding the pulse-induced crater, forming a symmetric, round contour centered on the pulse site. This feature may reflect a gradient of dissipated heat during the voltage pulse. In this context, Joule heating may play a dual role: enhancing Te depletion near the crater center and promoting adatom mobility across the surface. 

In summary, we demonstrate the formation of Bi(111) BL nanostructures on Bi$_2$Te$_3$(111) surfaces using voltage pulses from an STM tip. Positive pulses of $\ge +3$ V reproducibly induce pit craters surrounded by Bi BL islands. STM topography and FFT confirm that these islands retain the lattice constant and symmetry of the underlying Bi$_2$Te$_3$. We attribute the restructuring to a combination of electric field evaporation and local Joule heating. This method offers a new approach for constructing Bi BL nanostructures with site specificity, providing opportunities for the atomically precise patterning of layered topological materials using STM-based techniques.

\textbf{\textit{Acknowledgment.}} \textit{This work was supported by the U.S.\ Department of Energy, Office of Basic Energy Sciences, under Award No.\ DE-SC0016379.}
 
\bibliography{bibliography}

\clearpage
\appendix

\renewcommand{\thefigure}{SM\arabic{figure}}  
\setcounter{figure}{0}                       

\renewcommand{\thesection}{SM\arabic{section}}
\setcounter{section}{0}       

\section*{Supplementary Information}
\begin{figure}[h!]
\centering
\includegraphics[width=0.9\textwidth]{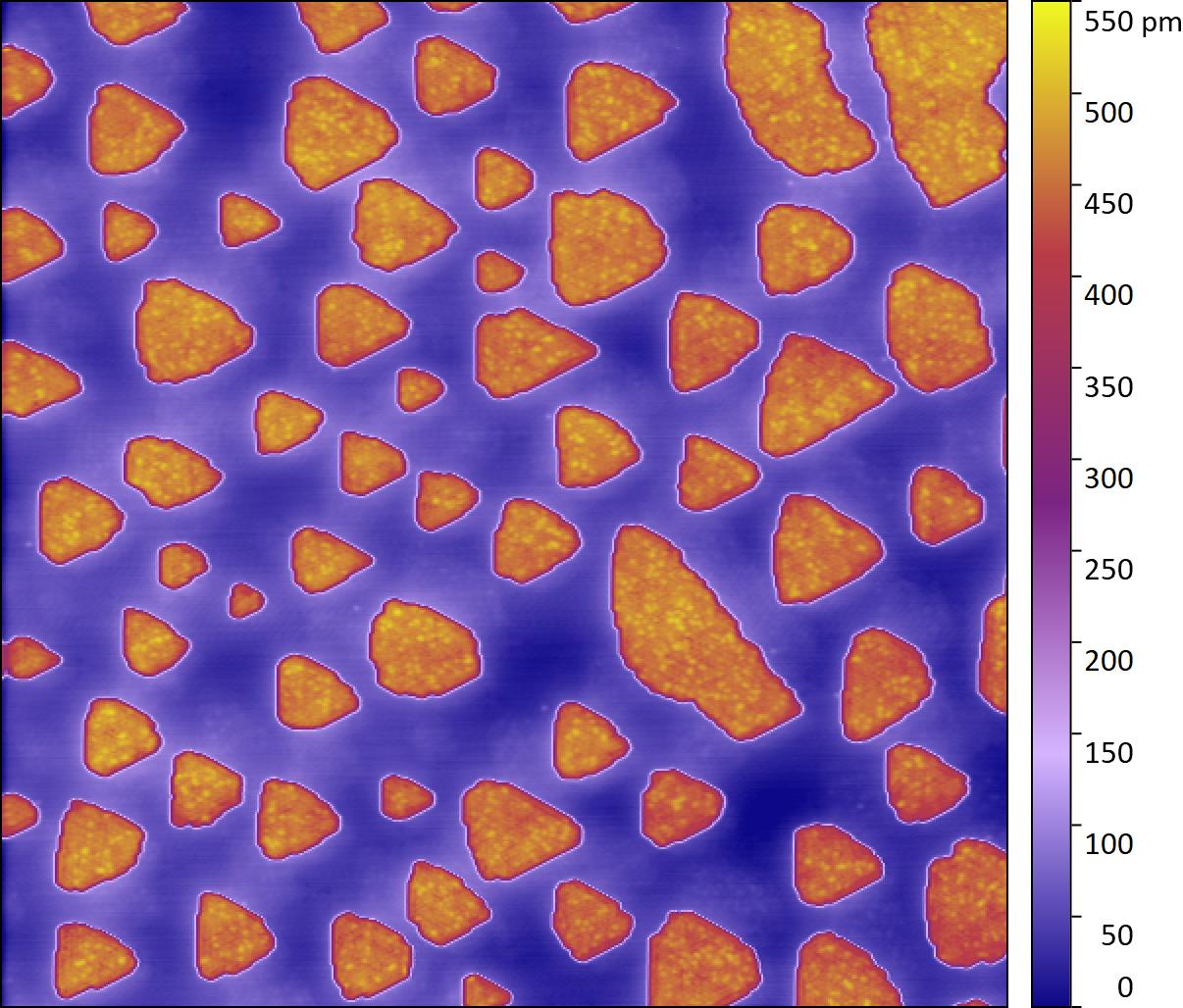}
\caption{\textbf{$150 \, \text{nm} \times 150 \, \text{nm}$ STM topographic image acquired at $+0.5 \, \text{V}$.} 
The mean height of Bi(111) BL islands was obtained by subtracting the mean Bi$_2$Te$_3$ terrace height from the mean island-top height, giving $0.392 \pm 0.006 \,\text{nm}$.}
\label{sm-fig1}
\end{figure}

\clearpage
\begin{sidewaysfigure}[h!]
\centering
\includegraphics[width=1.0\textwidth]{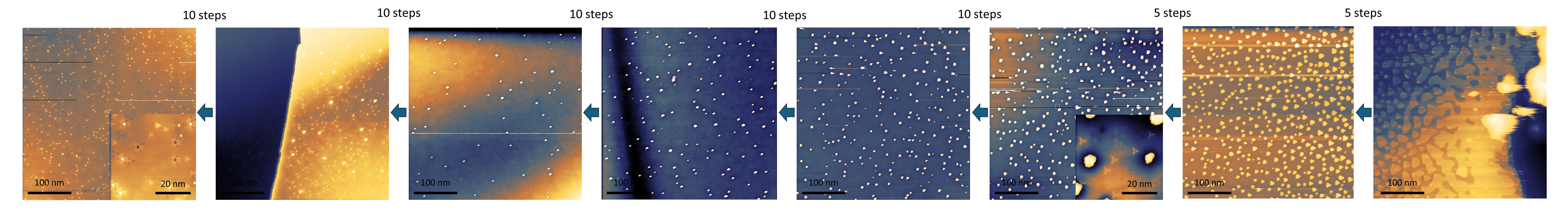}
\caption{\textbf{STM topographic series showing the evolution of the Bi$_2$Te$_3$ surface from the crater region (right) to the unmodified surface (left).} The rightmost image corresponds to the crater site, where coarse step terraces are observed. Moving leftward, the number of coarse steps increases, transitioning from areas with high-density Bi BL nanoislands (5–10 coarse steps between each image) to more sparsely distributed adatoms and vacancies. The leftmost image (60 coarse steps $\sim 2.46 \mu$m from the crater) marks the point where the Bi$_2$Te$_3$ surface returns to its native state, with vacancy distributions and Bi/Te adatoms consistent with pristine regions.}
\label{sm-fig2}
\end{sidewaysfigure}
\clearpage

\begin{figure}[h!]
\centering
\includegraphics[width=1\textwidth]{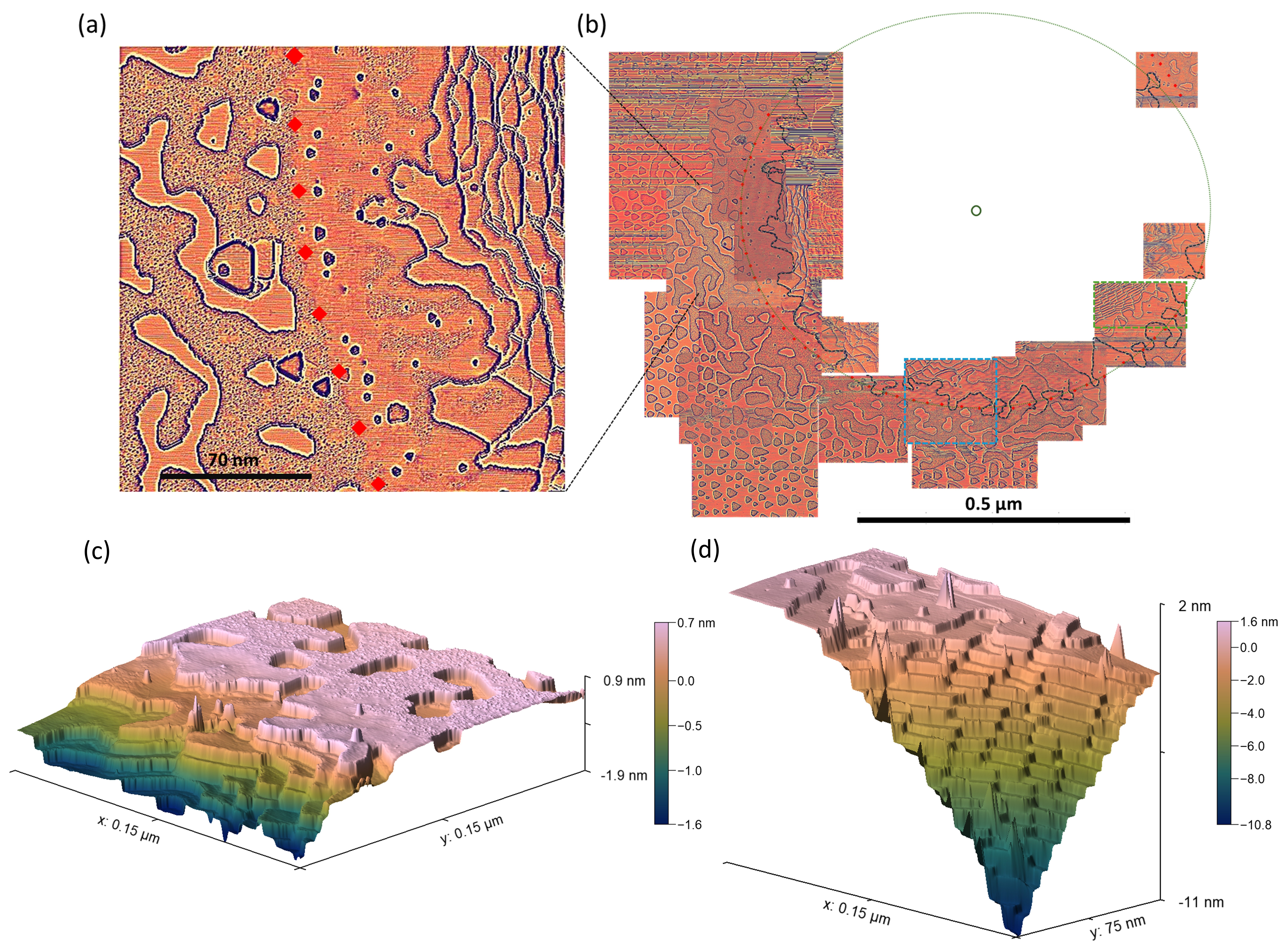}
\caption{\textbf{Laplacian-filtered and 3D-rendered STM topographic images of the $+6$V tip pulse.} 
\textbf{(a)} Close-up view showing the spatial evolution of native surface defects near the crater. [Red diamonds] mark the approximate boundary separating the defect-rich (left) and near defect-free (right) regions on a Bi BL surface. [Blue dashed line] outlines the edge of the topmost Bi BL terrace at the periphery. 
\textbf{(b)} Stitched Laplacian-filtered topographic images reassemble an overview of the $+6$V pulse region. [Green dashed ellipse] approximates the spatial boundary inferred from the [Red diamond] markers in (a), while the solid green circle indicates the estimated tunneling point during the voltage pulse.
\textbf{(c)} \& \textbf{(d)} Three-dimensional renderings of the areas outlined by the blue \& green dashed box in (b), respectively, reveal complex step height distributions (see further in Fig.~\ref{sm-fig4}).}
\label{sm-fig3}
\end{figure}

\begin{figure}[h!]
\centering
\includegraphics[width=1\textwidth]{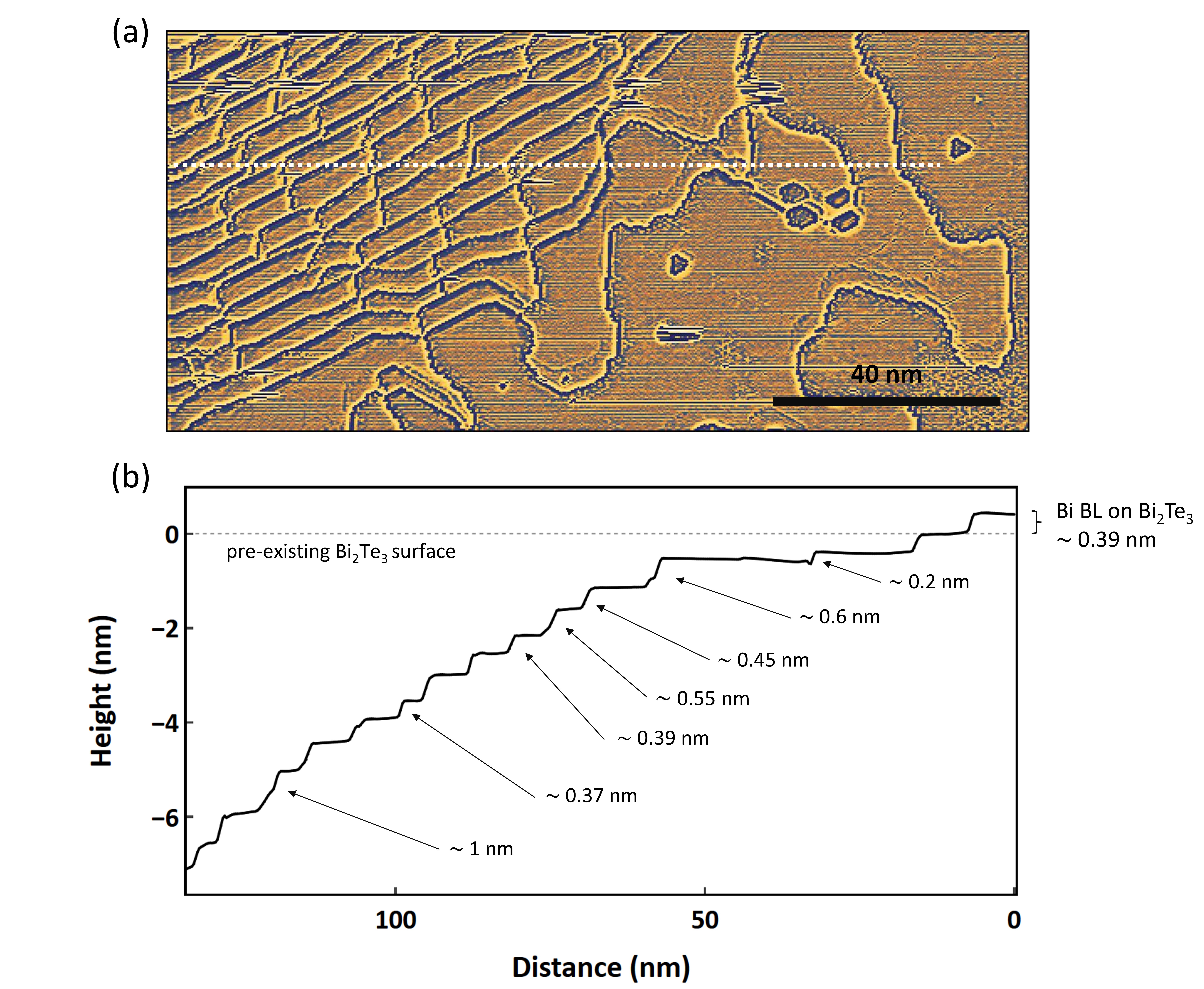}
\caption{
\textbf{(a)} Laplacian-filtered topograph of the area in Fig.~\ref{sm-fig3}(d).
\textbf{(b)} Height profile along the dashed trace in (a). In addition to the Bi BL step of $0.39$ nm, several sub-QL steps of $\sim\!0.20$, $0.45$, $0.55$, $0.60$, and $1.0$ nm are observed. We attribute these steps to intermediate stacking motifs of the mixed-stoichiometry Bi$_m$Te$_n$ homologous series, where Bi BLs are periodically inserted between Bi$_2$Te$_3$ QLs \cite{BiNTeM, BiNTeM2, Bi4Te3}. For example, a local sequence of three Bi$_2$Te$_3$ QLs adjacent to two Bi$_4$Te$_3$ blocks (a Bi BL inserted between every QL) would result in a $\sim 0.2$ nm step height.}
\label{sm-fig4}
\end{figure}
\end{document}